# Evaluating User Experiences in Mixed Reality

Dmitry Alexandrovsky*
dimi@uni-bremen.de
DMLab, University of Bremen
Bremen, Germany

Susanne Putze*
sputze@uni-bremen.de
DMLab, University of Bremen
Bremen, Germany

Valentin Schwind
Frankfurt University of Applied
Sciences
Frankfurt, Germany

Elisa D. Mekler
elisa.mekler@aalto.fi
Aalto University
Helsinki, Finnland

Jan David Smeddinck
jan.smeddinck@newcastle.ac.uk
Newcastle University
UK

Denise Kahl
denise.kahl@dfki.de
DFKI, Saarland Informatics Campus
Saarbrücken, Germany

Antonio Krüger
antonio.krueger@dfki.de
DFKI, Saarland Informatics Campus
Saarbrücken, Germany

Rainer Malaka
malaka@tzi.de
DMLab, University of Bremen
Germany

## 1 INTRODUCTION

Recent advances of Mixed Reality (MR) technology have enabled new research methods and interventions across various fields and allow for the design of highly immersive user experiences. By this, Virtual Reality (VR) and Augmented Reality (AR) research have become central topics in HCI. To measure these experiences, researchers apply a wide range of research methods using objective or subjective metrics [2]. Objective measures include behavioural metrics (e.g., gaze direction, movement amplitude), physiological measures, (e.g., EEG, EDA, ECG), and performance measures (e.g., time logging, success rates). Subjective self-reports through standardized or custom questionnaires remain a widely applied method for administering mid- and post-experience measures, such as the sense of presence [30] or being embodied using virtual avatars [29]. Alternatively, VR offers a wide range of opportunities for non-obstructive assessment methods of user experience, like objective measurements using biosignals [26, 27], or behavioural measures [32, 36]. Many of these measurement methods were adapted from use-cases outside of MR, in which interactions are often less immersive, and their validity of usage in MR experiments has not yet been validated. However, researchers are faced with various challenges and design alternatives when measuring immersive experiences. These challenges become even more diverse when running out-of the lab studies [20, 39]. Measurement methods for VR experience received recently much attention and research has already started to embed questionnaires in the Virtual Environment (VE) for various applications (e.g.,[14, 23]) as this allows to stay closer to the ongoing experience while filling out the survey [2, 7, 12, 27, 30]. However, there is a diversity in the interaction methods and practices on how the assessment procedure is conducted. This diversity in methods shows that there is no shared agreement on standardized methods of assessing the experience of being in the VR. Moreover, research pointed towards a multitude of open questions around methodological [2, 30], technical [26], social [41], and other challenges that require a focused investigation [20]. It appears crucial to work towards a shared agreement on assessment methods of VR user studies as researchers in the HCI community have to be aware of biases that may exist for their research methods of choice. AR research strongly orients on the research methods from VR, e.g., using the same type of subjective questionnaires. However, there are some crucial technical differences that require deliberate considerations during the evaluation. In this workshop, we exchange experiences with research methods in MR (i.e., AR/VR) user studies and examine the particular challenges of the different research methods. By this, our workshop launches a discussion of research methods which should lead towards standardizing assessment methods in MR user studies. The outcomes of the workshop will be aggregated into a collective special issue journal article.

## 2 BACKGROUND

Due to its immersive nature and a wide variety in technical setups, MR requires careful deliberation of the assessment methods when aiming to conduct immersive studies with human subjects. While MR allows for the implementation of diverse research settings, the technology itself affects the research results [40]. Research tries to counteract the disengaging and tedious qualities of (VR) user studies by making the tasks more appealing [42, 43]. The assessment of User Experience (UX) falls into two categories of *subjective* and

---

*Both authors contributed equally to this research.

*objective* metrics [24]. Most research attributes a sense of presence [32] and immersion as the central characteristic of UX in VR. There is a variety of standardized questionnaires to assess the presence, c.f., [30]. The major advantage of questionnaires is that they are easy to administer and generally don't require modifications of the VE [32]. However, post-experience questionnaires are not sensitive to state changes during the ongoing experience [16, 34]. Moreover, the existing scales (on presence) are often long and the items are not always fit well to the experiences. Further, it remains open for discussion if presence is actually a good candidate to describe the quality of a VR experience since a) it is difficult to measure and b) its relationship with user performance [15, 19, 44] or the fidelity [4, 35, 44] of the environment is ambiguous. Particularly while looking at applications in the mixed reality using a construct such as presence requires critical discussion. Yet, post-experience presence questionnaires remain the predominant method applied in the literature [32, 36]. Surveying UX within the VR experience received recent attention in the literature. Schwind et al. [30] contrasted the screen-based questionnaires against VR-embedded questionnaires and found that with embedded assessment the subjective responses in VR are more consistent. In contrast, others have shown that in-VR questionnaires may lead to inconsistencies [11]. To counteract for such inconsistencies, Alexandrovsky et al. [2] presented important usability criteria for in-VR questionnaires. Other tools that allow administering questionnaires in VR are the VR Questionnaire Toolkit [7], VRate [28]. Similarly, MRAT [21] is a toolkit for AR studies. These tools aim for a less-disruptive study flow and target problems of context-dependent forgetting [1, 10] due to environment change [25] which may bias responses.

Several approaches have been proposed for behavioral measures of UX, including gaze direction [22] responses to social [38], or threatening events [31], perception of discrepancy between VR and the physical space [37], or magnitude of postural responses [8]. Skarbez et al. point out that behavioral measures are objective, contemporaneous and non-intrusive and thus, they overcome some of the shortcomings of the subjective measures. However, in order to trigger specific behavioral responses the VE or evaluation procedure of the ongoing study requires specific manipulations, which are not always applicable [32]. Highly immersive experiences are expected to facilitate specific reaction patterns from the autonomous nervous system [6]. Physiological responses provide information about specific episodes of the experience [3, 16, 18, 26] and allow a better interpretation of subjective ratings and task performance [5]. However, these physiological signals are challenging to administer in MR scenarios. For example, assessing brain activity using Electroencephalography (EEG) sensor with Head Mounted Displays (HMDs) is cumbersome for both participants and researcher, as they may be uncomfortable to wear together and the electrical signals from the HMD can interfere with the EEG sensors [26]. Although research has shown that physiological measures are well applicable, Slater and Steed argue that physiological measures of presence can only be applied in anxious scenarios (e.g., a response to a threat) but that they are ineffective in mundane situations [33]. While measuring VR experience in the lab is diverse, measuring becomes even more technically and methodology challenging when running out-of the lab studies. Out-of the lab VR studies allow for larger variations in the settings [20] and require researchers for complex technical solutions [38, 39]. Ma et al. investigated how to enable telemetric web VR studies and to address the technical obstacles [17].

While AR research strongly orients on the research methods from VR (i.e., presence as a quality outcome of an experience), there are some crucial differences that require deliberate considerations. Especially optical see-through AR includes a high degree of interaction with the physical reality. Therefore, a strong focus on AR content might be disturbing [13] and a balanced fusion of reality with the virtual information is desired which should be ideally indistinguishable for the users. Therefore, measurement methods of immersive technology should account for both AR and VR. While a significant body of work developed standardized scales for measuring presence in VR (c.f., [32]), little research has been done on the development and adoption of the questionnaires for AR experiences. Georgiou and Kyza [9] developed the Augmented Reality Immersion (ARI) questionnaire, which conceptualizes immersion in AR applications on the three levels of *engagement*, *engrossment* and *total immersion* including subscales of interest, usability, emotional attachment, attention, presence and flow.

The presented literature outlines a series of challenges and possible pitfalls HCI faces in the context around measuring UX in immersive environments. Various toolkits and frameworks exists which address some of those challenges. However, there is still no agreement on assessing methods for UX in MR applications. This workshop targets at general and specific problems of UX research methods in MR and opens a critical discussion of existing research methods aiming to retain valid results when evaluating immersive technologies. The objectives of the workshops are finding a common ground of research practices and layout a research agenda towards standardized research methods of MR experiences.

## 3 ORGANIZERS

The organizers are all experienced researchers in the area of MR, evaluation of immersive experience, and the development of research methods. The co-organizers bring multiple perspectives from computer science, interaction design, psychology, and user engagement.

**Dmitry Alexandrovky** is a final-year doctoral student at the Digital Media Lab, University of Bremen, Germany. His research interests are immersive interaction, user engagement, and game design research. He works on interface designs for questionnaires in VR and developed an in-VR questionnaire toolkit. His research was awarded with 'Honorable Mentions' at CHI PLAY conferences.
**Susanne Putze** is a final-year doctoral student at the Digital Media Lab at the University of Bremen. Her research interests are in HCI, improvement of research workflows, and research communication methods. She works on measuring VR experiences using subjective questionnaires and physiological signals.
**Valentin Schwind** is professor for human-computer interaction at the Frankfurt University of Applied Sciences. His work explores immersive and multimodal user experiences in virtual and augmented reality. He is expert in research of quantifying immersion and presence. Valentin has received multiple awards at CHI and other HCI conferences for his research of avatars and virtual characters.

**Elisa D. Mekler** is an assistant professor at the Aalto University Department of Computer Science. Her research interests include the applications of psychological theories and methods in HCI, as well as the development and validation of UX questionnaires. Elisa's work has garnered multiple awards at CHI and CHI PLAY.

**Jan David Smeddinck** is an assistant professor at Open Lab and the School of Computing at Newcastle University in the UK. Building on his background in interaction design, serious games, web technologies, human computing, machine learning, and visual effects, his research interests include virtual-, mixed- and augmented-reality with a focus on applications in digital health and education.

**Denise Kahl** is a doctoral student at the German Research Center for Artificial Intelligence (DFKI). In her research she explores the relationship between virtual objects and their physical representations for tangible interaction in optical see-through Augmented Reality. She evaluates AR visualizations by measuring presence using subjective questionnaires.

**Antonio Krüger** is the CEO of the German Research Center for Artificial Intelligence (DFKI) and a professor of computer science at Saarland University heading the Ubiquitous Media Technology Lab (UMTL). He is an internationally renowned expert on human-machine interaction and artificial intelligence. His research focuses on Mobile and Ubiquitous Spatial Assistance Systems, combining the research areas Intelligent User Interfaces, User Modeling, Cognitive Sciences and Ubiquitous Computing.

**Rainer Malaka** is professor for Digital Media at the University of Bremen. He is managing Director of the Center for Computing Technologies (Technologiezentrums Informatik und Informationstechnik, TZI) and Director of the PhD program Empowering Digital Media that is funded by the Klaus Tschira foundation. His research focus is on multimodal interaction in MR, language understanding, entertainment computing, and artificial intelligence. Rainer is councillor of IFIP (International Federation for Information Processing) and chair of IFIP's technical committee on Entertainment Computing. He has an extensive experience in VR research and evaluation of VR applications from various research projects, including H2020s "first stage".

## 4 WEBSITE

To advertise the workshop, we will make our workshop website (http://evaluating-mr-ws.com/[1]) available upon the workshop acceptance, which features organizational aspects such as a Call for Participation, information about organizers, paper submission instructions and a workshop agenda, as well as later all contributions, presentations, and discussion outcomes (included the annotated Miro boards) of the workshop.

## 5 PRE-WORKSHOP PLANS

We plan to broadly advertise our Call for Participation via distribution lists and on social media (e.g., Twitter, Facebook). Meanwhile, we will also send personal invitations to potential researchers and practitioners from our network in the research community. The submission of workshop papers will be handled through a conference management system. All submitted workshop papers will be reviewed and selected by the workshop organizers (juried selection).

---

[1] URL will change after acceptance

We will share accepted workshop papers with the participants in advance of the conference. Participants are encouraged to publish a pre-print of their work, e.g., on arXiv or OSF.

## 6 WORKSHOP STRUCTURE

The workshop is planned as a single day workshop. The schedule consists of a mixture of prerecorded talks by the participants as well as active discussions and a breakout session. We expect around 15-20 participants, where 20 is the maximum. This group size allows on the one side for a versatile perspective on research methods and their challenges, and on the other side it enables intensive discussions with active participation of all participants. Preliminary schedule (CET):

**Welcome** (15:00 – 15:15): Opening presentation to outline the workshop motivation and goals.

**Paper session 1** (15:15 – 16:30): Current challenges and barriers of measuring UX in MR

**Coffee break** (10 min)

**Paper session 2** (16:40 – 17:55): Future directions of measuring methods for UX in MR We will have two paper sessions with max. 10 participants in each session. Each paper session is split up into two blocks of about five papers. In the blocks we will show the pre-recorded video presentations, including a short introduction of the presenter. The blocks end with an open discussion on the session's topic. To engage the participants into the discussion, we will prepare ice breaking questions.

**Coffee break** (15 min)

**Breakout session** (18:10 – 18:50): In the breakout session, all participants will discuss in small groups of 3-4 people for 20 min. The groups will be assigned in advance according to the paper topics. During the breakout session, the participants should brainstorm and aggregate their discussions in mind maps and charts. After that, the groups will present their outcomes to the workshop and open the discussion.

**Closing and wrap up** (18:50 – 19:00): Workshop results, including best practices, and experiences from the field trip will be documented. Remaining open questions will be wrapped up, follow-up activities will be discussed.

We will end the workshop day with a virtual social event in a on Altspace, e.g., a basketball tournament. To facilitate the discussion in the breakout session and on the paper presentation videos as well as to capture their outcomes we will deploy collaborative online Miro boards which allows for collaborative discussions remotely, and persist discussion results.

## 7 DISTANCE ENGAGEMENT

To incorporate both participants remotely, we will base the meeting on an online platform (Zoom or in line with the CHI 2021 centralized communication system) which allows for showing pre-recorded presentation videos including presenter introductions, discussions as well as breakout sessions for a subset of the participants. To ensure accessibility, we asked all participants to pre-record and subtitle their presentation, and upload them before the workshop. This allows all participants to watch the presentations regardless of bad connectivity. To support the discussion in the breakout session and on the papers, we will deploy collaborative Miro boards which allows for collaborative discussions remotely. For interactive but distance socializing we will organize a virtual get-together at the end of the workshop day. Further, we will provide a chat platform (Slack, or other platforms in line with the CHI 2021 centralized communication system) for communications during coffee breaks, as well as before and after the workshop.

## 8 POST-WORKSHOP PLANS & EXPECTED OUTCOMES

We expect our workshop to

- Connect a community of researchers and practitioners in investigating the challenges and opportunities of MR measurement and evaluation methods of UX.
- Identify current challenges and barriers of MR evaluation methods.
- Outline guidance for research methods and interaction design for MR user studies.

To ensure these outcomes of the workshop, we will:

- Summarize the workshop outcomes and share all presented materials on the workshop website.
- The outcomes will be aggregated on the workshop website.
- Disseminate the workshop results to the Human-Computer Interaction (HCI) community in form of a collective journal article co-authored by the organizers and the participants.

## 9 CALL FOR PARTICIPATION

To measure MR experiences, researchers apply a range of research methods using objective (e.g., biosignals, logging, behavioural), and subjective (e.g., questionnaires) metrics. However, the assessment methods are heterogeneous and miss consistency among the user studies which impedes transferability of the results.

This one-day virtual CHI2021 workshop will focus on common practices of evaluation methods and their methodological, technical, and design challenges. We invite researchers and practitioners from all subfields of HCI to drive the research agenda of the research practices, technologies, and challenges of MR user studies. This workshop invites submissions of position papers (2-4 pages excluding references according to the (single column) ACM Templates), covering but not limited to the following topics:

- Measurement methods (behavioural, objective, subjective) for single- or multi-user MR
- Technical challenges/solutions/artifacts for assessment methods in and outside the lab. E.g., interaction for in-VR measurements, use of biosignals, assessing behavioral measures
- Experimenter-participant communication (e.g., telepresence, avatarization)

Submissions will be selected by the workshop organizers based on the relevance to the workshop topic and their potential to engender insightful discussions at the workshop. At the workshop, accepted papers will have a 3-4 minutes video presentation. At least one author of the accepted paper must attend the virtual workshop. All participants must register for both the workshop and for at least one day of the conference.

**Important Dates**

- Submission Deadline: February 21st, 2021 at 12pm PT
- Notification: TBA
- Workshop Day: May 7t/h8th/9th, 2021, virtual

For submission and further information, please visit: http://evaluating-mr-ws.com/